\newcommand{\text}{\rm} 
\def\Tr{{\rm Tr}\,} 
 
\def\Eq#1{Eq.~(\ref{#1})}

\documentclass{PoS} 
\title{Strings in SU(N) gauge theories in 2+1 dimensions: beyond the fundamental representation} 
 
\ShortTitle{Strings in SU(N) gauge theories in 2+1 dimensions: beyond the fundamental representation} 
 
\author{\speaker{Barak Bringoltz}$^{ab}$ and Michael Teper${^a}$\\ 
        \llap{$^a$}Rudolf Peierls Centre for Theoretical Physics\\ 
University of Oxford\\ 
1 Keble Road, Oxford, OX1 3NP\\ 
        \llap{$^b$}Isaac Newton Institute for Mathematical Sciences\\ 
University of Cambridge\\ 
20 Clarkson Road, Cambridge, CB3 0EH, U.K.\\ 
E-mail: \email{barak@thphys.ox.ac.uk}, \email{m.teper1@physics.ox.ac.uk}}

\abstract{We calculate energies and tensions of closed $k$-strings 
in (2+1)-dimensional SU(N) gauge theories with N=4,5,6,8. When we 
study the dependence of the ground state energy on the 
string length, we find that it is well described by a Nambu-Goto (NG)
free bosonic string for large lengths. At shorter lengths 
we see deviations which we fit, and this allows us to control the 
systematic error involved in extracting the tension. We compare the resulting string tensions with Casimir scaling, which we find
to be lower than our 
data by $1\%-4\%$. Extrapolating our results to $N=\infty$ we see 
that our data fits more naturally to $1/N$ rather than $1/N^2$ 
corrections. Finally, we see that the full spectrum of the $k$-string states falls into sectors that belong to particular irreducible representations of $SU(N)$.} 
 
\FullConference{The XXV International Symposium on Lattice Field Theory\\ 
                 July 30 - August 4 2007\\ 
                 Regensburg, Germany} 
 
\begin{document} 
 
\vspace{-1cm}
\section{Introduction} 
\label{intro} 
\vspace{-0.35cm}

In this work we study $SU(N)$ gauge theories in $D=2+1$ space-time dimensions and focus on the energies and tensions of closed strings that carry flux in $SU(N)$ representations whose N-ality $k$ is larger than one. (For references on related lattice works see our companion contribution to these proceedings \cite{NGpaps} as well as \cite{kstrings}-\cite{Gliozzi}).
The reason we restrict ourselves to $D=2+1$ dimensions is closely related to our motivation in the related study \cite{KN_papers}, where we tested the Karabali-Kim-Nair (KKN) analytic prediction \cite{KKN} for the fundamental string tension $\sigma$. As a natural continuation to that work, we now aim to test how accurate is the following KKN prediction for the tension of a string in a general $SU(N)$ representation ${\cal R}$.
\vspace{-0.3cm}
\begin{equation} 
\sigma_{\cal R} = \sigma \cdot \frac{C_{\cal R}}{C_{1}}. 
\label{KKN_casimir} 
\end{equation} 
Here $C_{\cal R}$ is the quadratic Casimir of the representation, 
and $C_1=(N^2-1)/2N$ is the Casimir of the fundamental representation. We note here that the above `Casimir scaling' of string tensions is also predicted by other 
approaches to $2+1$ dimensions (for example see \cite{Greensite}). The KKN prediction lacks the physics of screening, and we thus take a practical point of view and regard \Eq{KKN_casimir} as an approximate prediction for the tensions of stable $k$-strings, and for the asymptotic energy per unit length of excited $k$-strings states with flux in an excited representation.

Besides comparing to \Eq{KKN_casimir}, we are also interested in the way the $k$-string energy depends on the string length $l$ (which will reveal its central charge), on the manner in which the planar limit is approached (i.e.  whether the corrections to $N=\infty$ scale like $1/N$ or $1/N^2$), and on a curious pattern of degeneracies seen in previous studies of the $k$-string spectrum.

\vspace{-0.5cm}
\section{Methodology} 
\vspace{-0.3cm}

We define the gauge theory on a discretized periodic Euclidean 
three dimensional space-time lattice, with spacing $a$ and, 
typically, with $L^2_s L_t$ sites. The action we use is the ordinary Wilson action, where the bare coupling $\beta$ is 
related to the dimensionful coupling $g^2$ by $\lim_{a\to 0}\beta=\frac{2N}{ag^2}.$
In the large--$N$ limit, the 't Hooft coupling $\lambda=g^2N$ is 
kept fixed, and so we must scale $\beta=2N^2/\lambda \propto N^2$ 
in order to keep the lattice spacing fixed (up to $O(1/N^2)$ 
corrections). We calculate observables 
by performing Monte-Carlo simulations of the Euclidean path integral, in which we use a 
mixture of Kennedy-Pendelton heat bath and over-relaxation steps 
for all the $SU(2)$ subgroups of $SU(N)$.

We measure the energy of flux tubes closed around a spatial torus, from the 
correlators of suitably smeared Polyakov loops that have vanishing transverse 
momentum \cite{Teper_Nd3,LTnLTW}. For each Hilbert space sector of given N-ality $k$ we construct lattice operators 
that couple to states of that N-ality. These are given by 
$\Tr U^k,\,\Tr U^{k-1}\Tr U\,\dots,\,\left(\Tr U\right)^{k}$, 
where $U$ is the path-ordered product of smeared links around the spatial torus. 
We then construct the full 
correlation matrix and use it to obtain best estimates for the 
string states using a variational method applied to the transfer 
matrix $\hat{T}=e^{-aH}$ (see for example \cite{Teper_Nd3} and references therein).

Our study is logically divided into two.
We first investigate the way the $k$-string energy $E$ depends on its length $l$. In \cite{KN_papers, NGpaps} we have discussed the theoretical possibilities for $E(l)$, and we will not reiterate that discussion here, but rather just quote its conclusion : a natural way to fit our data for the energy is with 
\vspace{-0.3cm}
\begin{equation}
E^2_{k}(l) = E^2_{\rm NG} - \frac{C_k}{\left(l\sqrt{\sigma_k}\right)^3} \quad ; \quad E^2_{\rm NG} = \left(\sigma_k l\right)^2 -\sigma_k \frac{\pi}{3},\label{fit}
\end{equation}
where $E_{\rm NG}$ is the ground state energy of a closed string in the Nambu-Goto string theory.

For the $k=1$ case we found that our data is very well described by this ansatz with $C_1\stackrel{<}{_\sim}0.3$. 
We now ask whether this situation persists for $k>1$ strings as well. This will also tell us whether the $k$-strings belong to the same IR  universality class as the $k=1$ string. 
We perform these measurements for $SU(4)$ at $\beta=28.00,50.00$, $SU(5)$ at $\beta=80.00$, $SU(6)$ at $\beta=59.40,90.00$, and $SU(8)$ at $\beta=108.00,192.00$. These bare couplings correspond to lattice 
spacings of $a\simeq 0.06,0.08,0.11$ fm, depending on $N$. The string lengths $l$ ranged between 
$\sim 0.45$ fm and $\sim 3$ fm, again depending on the values of $N$ and  
$a$.\footnote{For more details on the lattice parameters of our field 
configurations see \cite{NGpaps}.}

After we obtain an estimate for $E_k(l)$ we use it to extract string tensions from string energies which were measured on a set of lattices with increasingly small spacings in the range $a\simeq 0.05-0.2$ fm. This is done only for strings whose length obeys $l\stackrel{>}{_\sim}3/\sqrt{\sigma}\simeq 1.4-1.5$ fm. This way of extracting tensions controls the systematic error involved in the usual neglect of the sub leading corrections to the Luscher term. 
Once we obtain the continuum string tensions, we extrapolate our results to the large-$N$ limit. This is particularly 
interesting since there exists a controversy in the literature with respect to the possibility of having $1/N$ corrections in the $k$-string tensions \cite{SA,MK}.

\vspace{-0.5cm}
\section{Results : length dependence of the $k$-string energies and their conformal anomaly} 
\label{results1} 
\begin{figure}[htb]
\centerline{ 
\includegraphics[width=7.25cm]{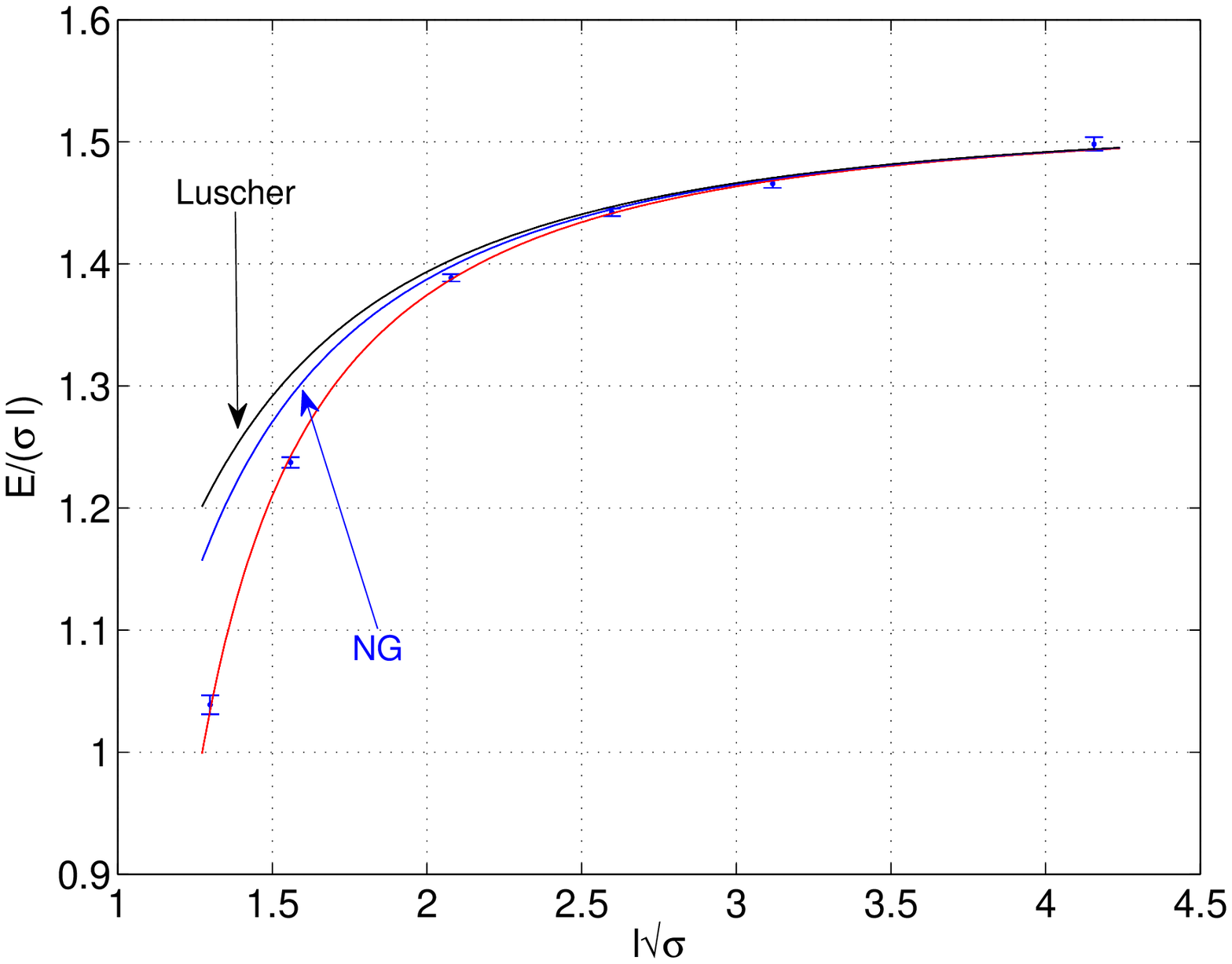} \quad \includegraphics[width=7cm,height=5.5cm]{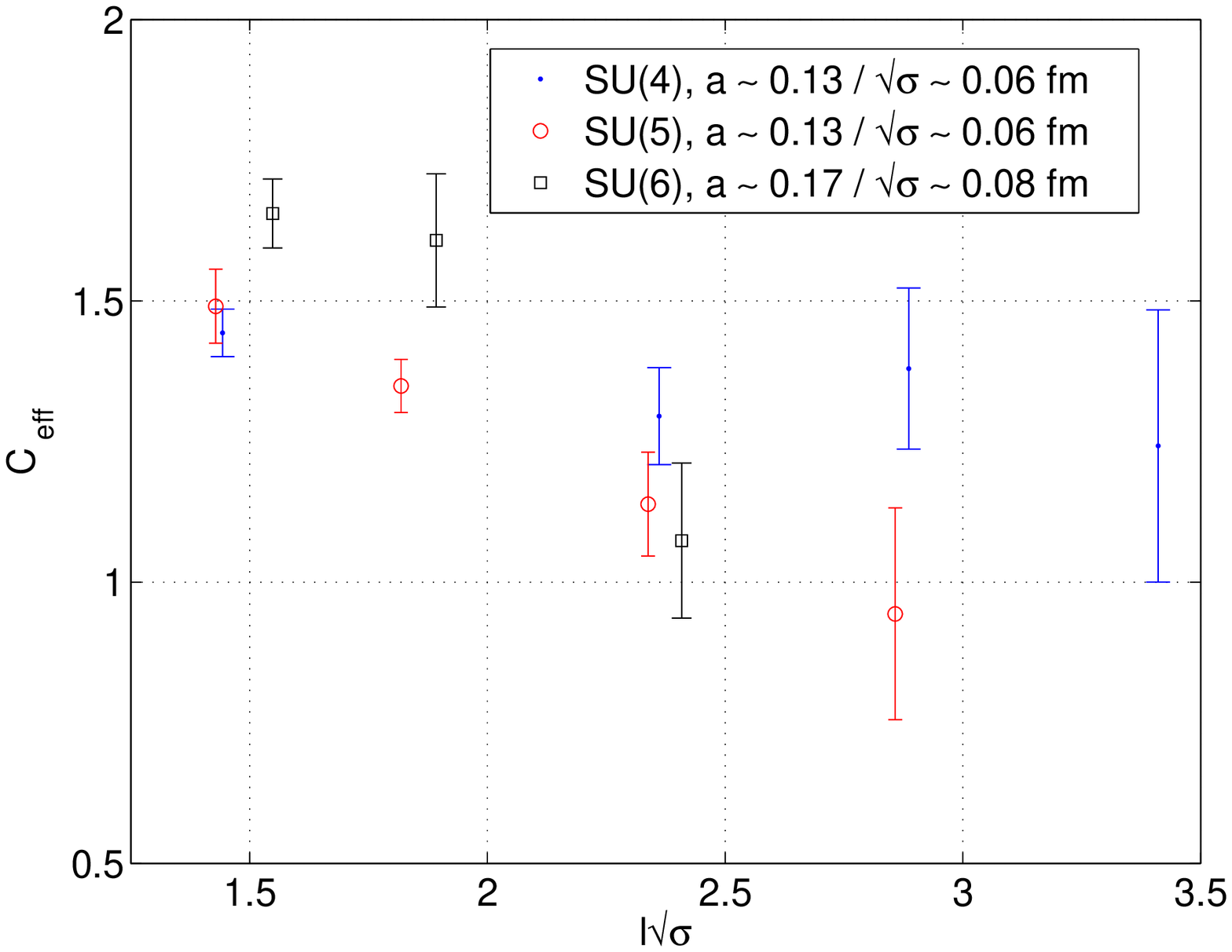} 
} 
\caption{\underline{Left:} Ground state energy of the $k=2$ string in $SU(5)$ and $\beta=80.00$. Our fit in red, and in blue(black) the NG(Luscher term) predictions. \underline{Right:} The value of $C_{\rm eff}$ (see text) of the $k=2$ strings in $SU(4,5,6)$ for $\beta=50.00,80.00,90.00$, respectively (lattice spacings presented in the legend).} 
\label{N5k2B80gs} 
\end{figure}

\vspace{-0.5cm}
 In the left panel of Fig.~\ref{N5k2B80gs} we present the energy of 
the $k=2$ string for $SU(5)$ at
$\beta=80.00$. The plot shows the energy divided 
by $\sigma l$ (here $\sigma$ is the fundamental string tension 
 which we obtain in \cite{NGpaps}) vs. the length 
in physical units, $l\sqrt{\sigma}$. The string tension in 
lattice units is $a^2\sigma = 0.016874(12)$ which gives a lattice 
spacing of $a\simeq 0.058$ fm, and tells us that our string length stretches from $\simeq 0.6$ fm to $\simeq 1.85$ fm. The red line 
that goes through our data is a fit of the form \Eq{fit} which 
results in the ratio $\sigma_2/\sigma=1.5244(21)$ and 
$C_2=1.41(7)$ (the fit is good with $\chi^2/dof\simeq 3/4$). The 
coefficient $C_2$ is thus much larger than the corresponding one for 
$k=1$, which was $0.0554(139)$ for this data \cite{NGpaps}. This reflects 
the fact that the NG prediction is a much better approximation for 
$k=1$ than it is for $k>1$, which can be easily seen by comparing the left panel of Fig.~\ref{N5k2B80gs} to the corresponding plot for $k=1$ \cite{NGpaps}. 

The results for all the gauge groups that we study are similar and can be encompassed in a single 
formula with $C_2=3(2)$. For higher values of $k$ the results are  less accurate, but we can still fit them and find that 
taking $C_3=4.5(2.0)$ and $C_4=5.5(1.5)$ for $k=3$ and $k=4$, 
respectively, describes all our data. In practice, provided that the 
lengths of our strings obey $l\stackrel{>}{_\sim}3/\sqrt{\sigma}$, we see that the correction term in 
\Eq{fit} is at most a $0.5\%$ contribution to the energy.
 
We now examine the universality class of the string by fitting pairs of adjacent points in the left panel of Fig.~\ref{N5k2B80gs}, and in the corresponding data sets for all other values of $N$ and $k$, with the form $E^2=\left(\sigma_k l\right)^2 - \sigma_k \frac{\pi}{3} \times C_{\rm eff} \label{fit_C}.$
As the points we fit have larger and larger values of $l$, then $C_{\rm eff}$ should approach the central charge of the $k$-string. In the case of $k=1$ we have very strong evidence that $C_{\rm eff}\stackrel{l\to \infty}{\to} 1$ (see \cite{KN_papers,NGpaps} and references within). In the $k>1$ the situation is harder to pin down and in addition there is a recent prediction \cite{Gliozzi} suggesting that $C_{\rm eff}\stackrel{l\to \infty}{\to}\sigma_k/\sigma$. 
We present our results in the right panel of Fig.~\ref{N5k2B80gs} for the cases $N=4,5,6,8$ and $k=2$, where it is reasonably clear that $C_{\rm eff}$ decreases towards $1$ as $l$ increases. We have similar results for $k=3,4$, which are, however, less accurate due to the larger energies. 

Let us now pause to make the following comment. The decrease of $C_{\rm eff}$ becomes clear only above $1$ fm, and its possible that this is the main cause for the difference between our conclusions and those of \cite{Gliozzi}, where the maximum string length was $\sim 0.9$ fm. 

\vspace{-0.5cm}
\section{Results : the string tensions in the continuum and the large-$N$ extrapolation} 
\label{results2} 
\begin{figure}[htb]
\centerline{
\includegraphics[width=7cm]{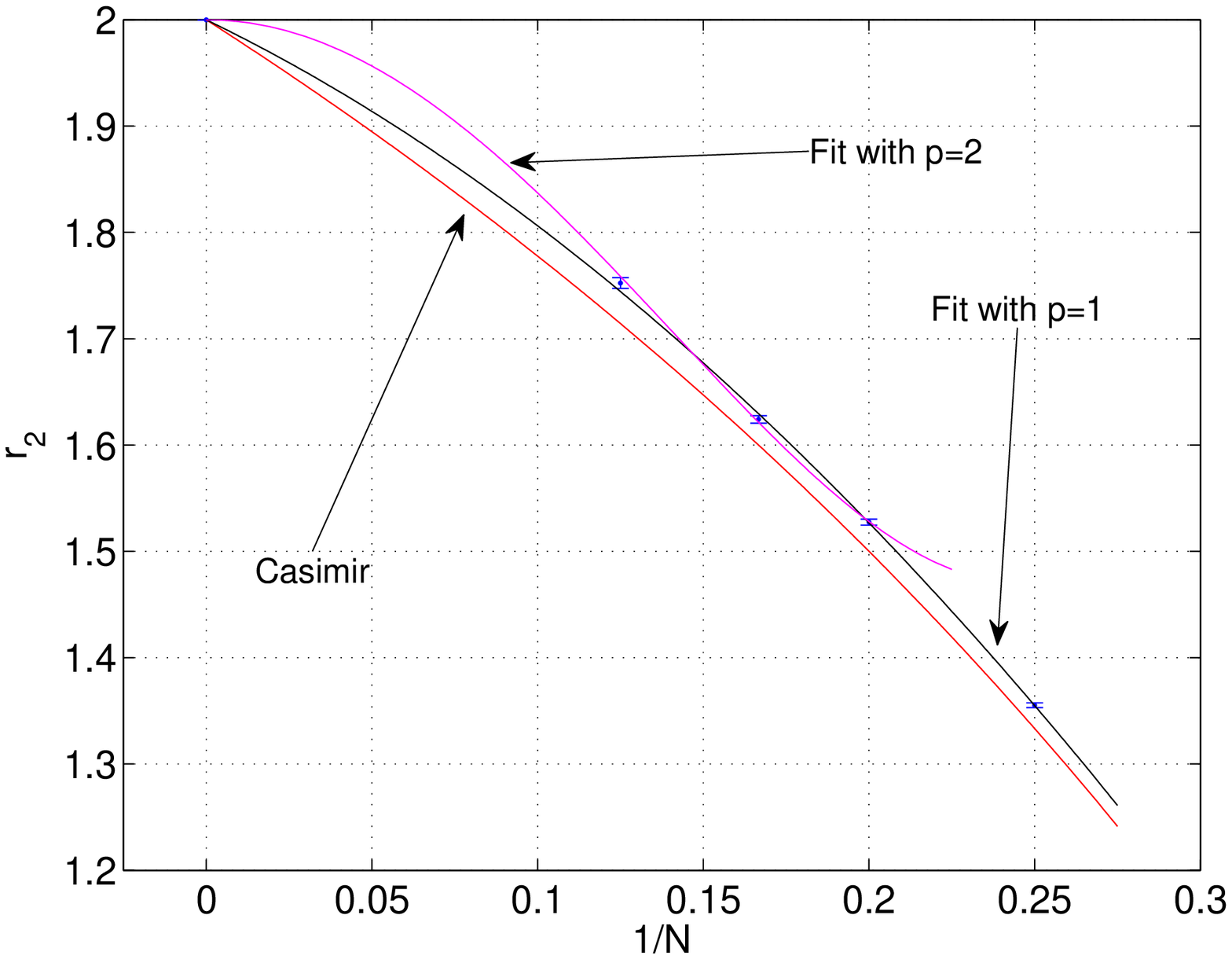} \qquad \includegraphics[width=7cm]{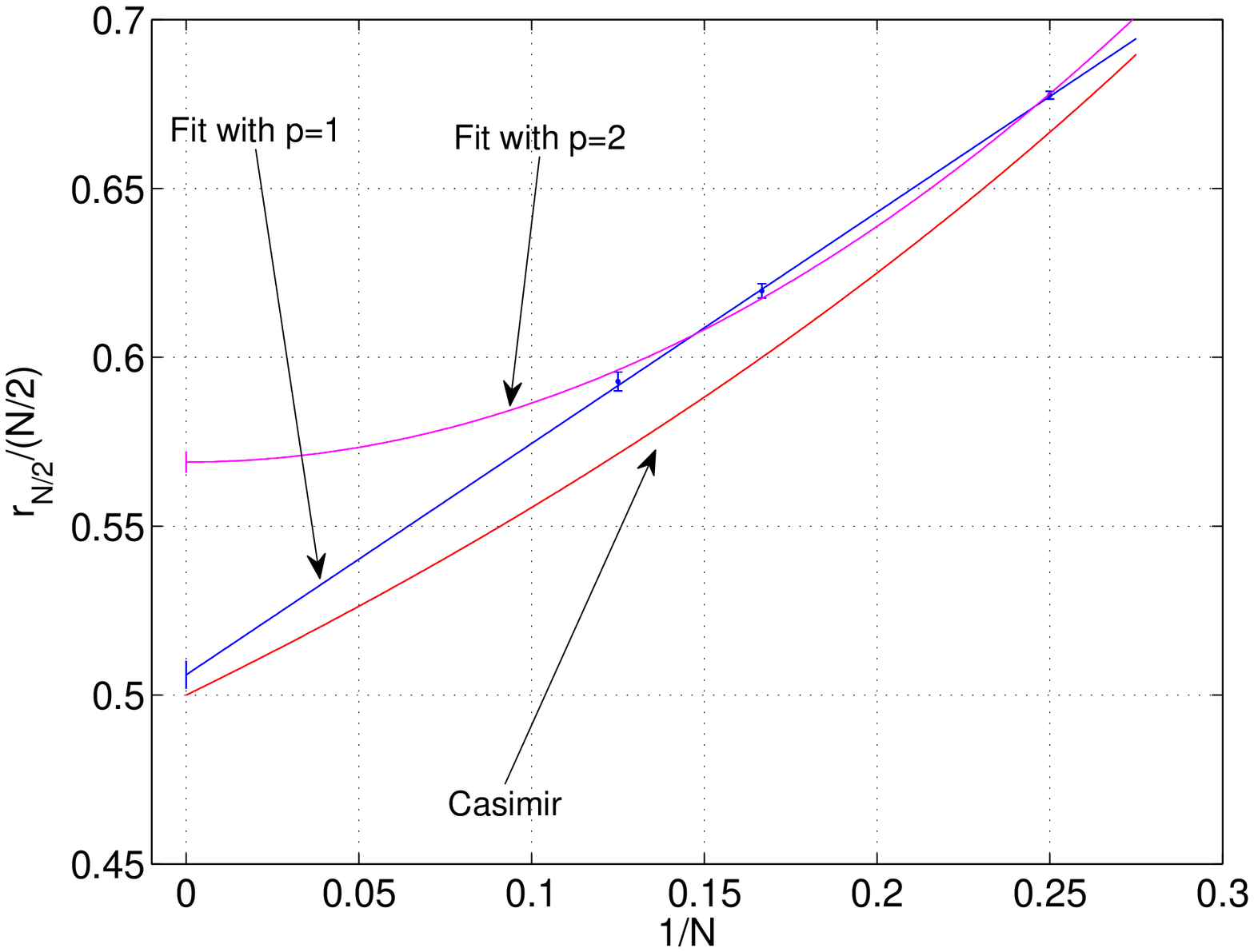}
}
\caption{The extrapolation of $k=2$ (left panel) and $k=N/2$ (right panel).}
\label{largeN}
\end{figure}

\vspace{-0.4cm}
 We now use the empirically determined \Eq{fit} 
to extract string 
tensions from string energies that we measure on a wide range of lattice spacings ranging between $a\simeq 0.2$ fm and 
$a\simeq 0.05$ fm. All the strings the we use have a length of at 
least $1.35$ fm. The extrapolation of the ratios $r_k\equiv \sigma_k/\sigma$ to the continuum, and its comparison to the Casimir scaling is shown in Table~\ref{cont_all}.\footnote{The determination of the fundamental tension in these cases is described in \cite{KN_papers}} All the continuum extrapolations had 
a acceptable $\chi^2/dof$ except for the $k=2$ of $SU(6)$ where we find that our data is too scattered to be well fit by a smooth fitting ansatz.
\begin{table}
\centerline{
\begin{tabular}{|c|c|c|c|c|c|c|c|}\hline
$r_k(N)$ & $r_2(4)$ & $r_2(5)$ & $r_2(6)$ & $r_2(8)$ & $r_3(6)$ & $r_3(8)$ & $r_4(8)$ \\ \hline
Lattice &  $1.3553(23)$ & $1.5275(26)$ & $1.6242(35)$ & $1.7524(51)$ & $1.8590(63)$ & $2.1742(187)$ & $2.3725(111)$ \\ \hline
Casimir & $1.3333\dots$ & $1.5$ & $1.6$ & $1.7142\dots$ & $1.8$ & $2.1429\dots$ & $2.2857\dots$ \\ \hline 
\end{tabular}
}
\caption{The continuum extrapolation of $r_k(N)\equiv \frac{\sigma_k}{\sigma}(N)$ and the comparison with Casimir scalings.}
\label{cont_all}
\end{table}
We proceed to perform two types of large-$N$ extrapolation. 
The first is for $k=2$ (left panel of Fig.~\ref{largeN}) and the second is for $k=N/2$ (right panel of the figure). In both cases we present the Casimir scaling prediction in red, and two type of fits, that either allow or exclude $1/N$ corrections.

We begin by extrapolating $r_{k=2}$ to $SU(\infty)$. 
Since at $N=\infty$ one expects the $r_2=2$ we use the ansatz
$r_2=2 - \frac{a}{N^p} - \frac{b}{N^{2p}}$ with $p=1,2$. For $p=1$ our fit gives $a=1.51(5)$ and $b=4.3(2)$, but a $\chi^2/dof\simeq 2.2$. This high value of $\chi^2$ comes from the data point of $SU(6)$ which, as mentioned above, suffers from a low confidence level. To check the sensitivity of the fit to this point, we  drop it  from the fit and find an acceptable $\chi^2$ with similar values for the fit parameters $a,b$. When $p=2$, however, we find no acceptable fit, and are led to drop the point with the lowest value of $N=4$. This results in $a=17.8(3)$ and $b=-150(9)$ and a still large $\chi^2/dof\simeq 2.1$. Also, the wavy behaviour of this fit (magenta line), suggests that the $p=2$ ansatz is questionable.

Proceeding to extrapolate $\sigma_{k=N/2}$ to $SU(\infty)$, we 
begin with a cautionary remark. In this extrapolation, we use the tension of the $k=4$ strings in $SU(8)$, which has a relatively large mass. This means that while its statistical error is also large, it may suffer from an even larger systematic error. Nonetheless, we proceed to fit our data with the ansatz $\frac{r_{N/2}}{N/2} = a + \frac{b}{N^p}$ with $p=1,2$. 
For $p=1$ we find that $a=0.506(4)$ and $b=0.68(2)$ is a very good fit with $\chi^2/dof\simeq 0.27$. Also it is interesting to note that $a$ is in fact consistent with the Casimir scaling prediction. In contrast, the $p=2$ best fit has a high $\chi^2/dof\simeq 2.7$ (with $a=0.569(3)$ and $b=1.75(5)$).

\vspace{-0.4cm}
\section{Near degeneracies in the $k$-string excited state spectrum}
\label{deg}
\begin{figure}[htb] 
\centerline{ 
\includegraphics[width=7cm,height=5.5cm]{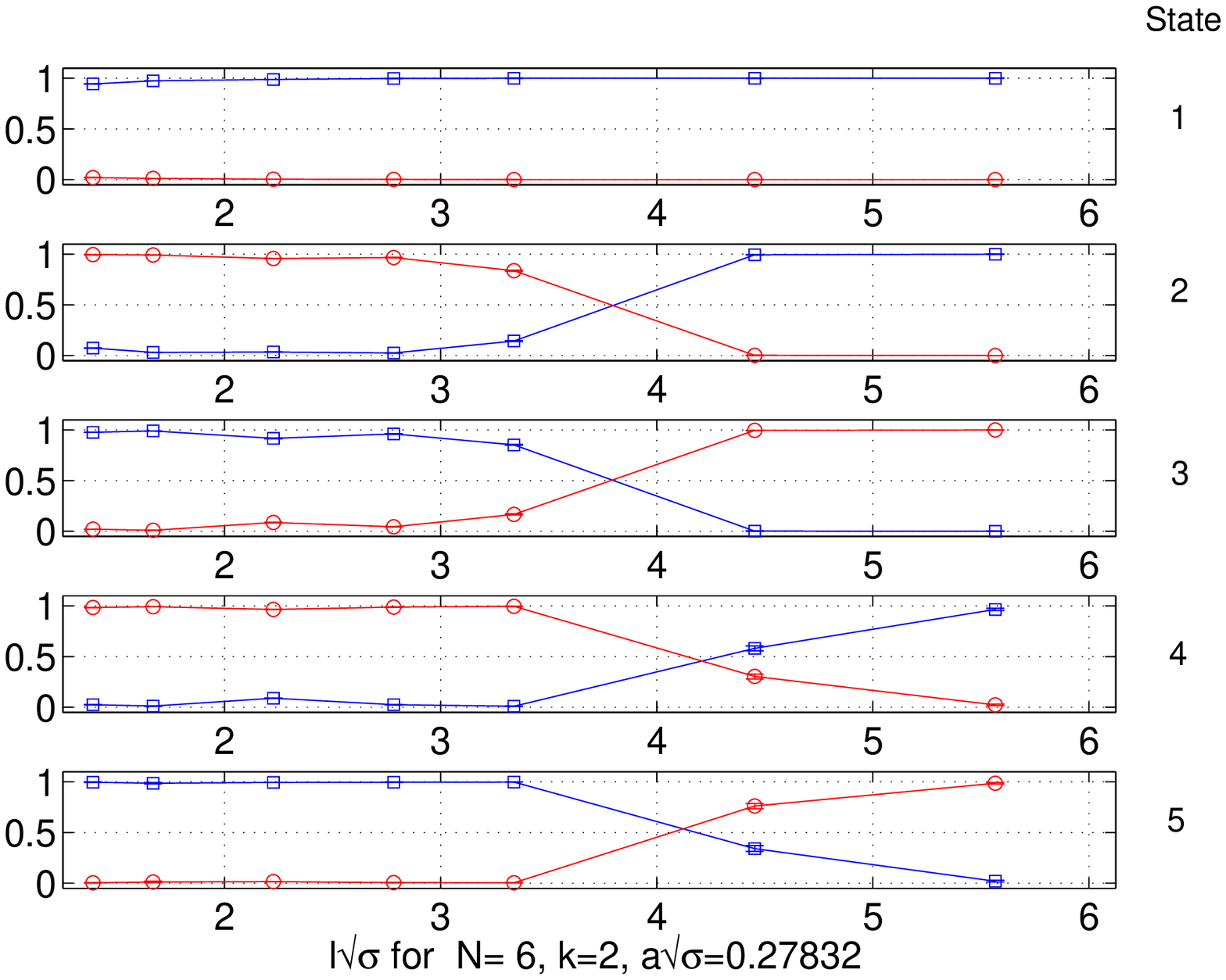} \qquad \includegraphics[width=7cm,height=5.5cm]{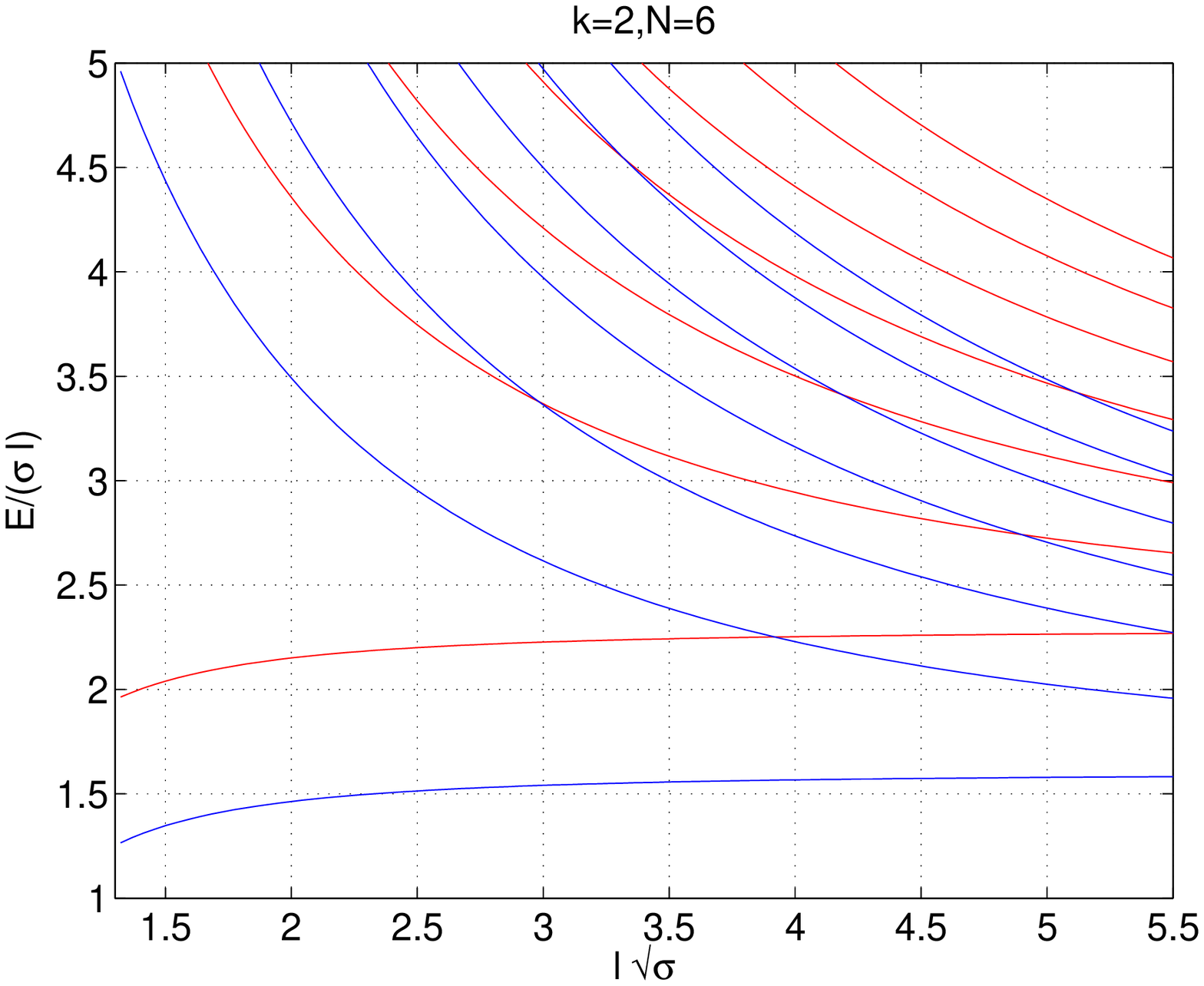} 
} \caption{\underline{Left:} Overlaps of the five lowest 
states in the $k=2$ spectrum of $SU(6)$ at $\beta=59.40$ ($a\simeq 
0.12$ fm) vs. the string length. In blue(red) are the overlaps 
onto the antisymmetric(symmetric) representation. 
\underline{Right:} The spectrum of energies vs. the string length 
for $SU(6)$ and $k=2$ of the $2$-NG model (see text). As in the 
left panel, blue(red) denotes the energies of the 
antisymmetric(symmetric) representation.} 
\label{over} 
\end{figure}

\vspace{-0.4cm}
We now use our data to revisit the issue of the near degeneracies in the $k$-string spectrum that were seen in \cite{LTnLTW,AL}, with the clear advantage that our new data contains measurements of the string spectrum for a 
variety of string lengths $l$.
We begin by 
focusing on the operators that couple best to the lowest states (as determined by our variational calculation), 
and calculate their overlap onto particular $SU(N)$ 
representation. We present the dependence of these overlaps on the 
string length for the five lowest states of the $k=2$ string in 
$SU(6)$ (left panel of Fig.~\ref{over}). This figure tells us that the ground 
state is always in the anti-symmetric representation, while the 
other states may change their `representation content'. In particular, the
first excited state is symmetric when the string is short, and becomes anti-symmetric for longer lengths. The opposite
happens for the second excited state, and a similar pattern is 
seen for the third and fourth excited states. 

To interpret these results we suggest the following simple model. Consider two non interacting NG free bosonic strings that carry fluxes in 
irreducible $SU(N)$ representations, and whose tensions scale
according to \Eq{KKN_casimir}. This model's spectrum for $SU(6)$ is presented in the right panel of Fig.~\ref{over},\footnote{For a discussion on the excited state spectum of the NG model see \cite{NGpaps}.} where we see that it works well in 
predicting the switching of states as well as the approximate $l$ at which this occur. We find that this model 
works also for other values of $N$ and $k$.

\begin{figure}[htb] 
\centerline{ 
\includegraphics[width=7cm]{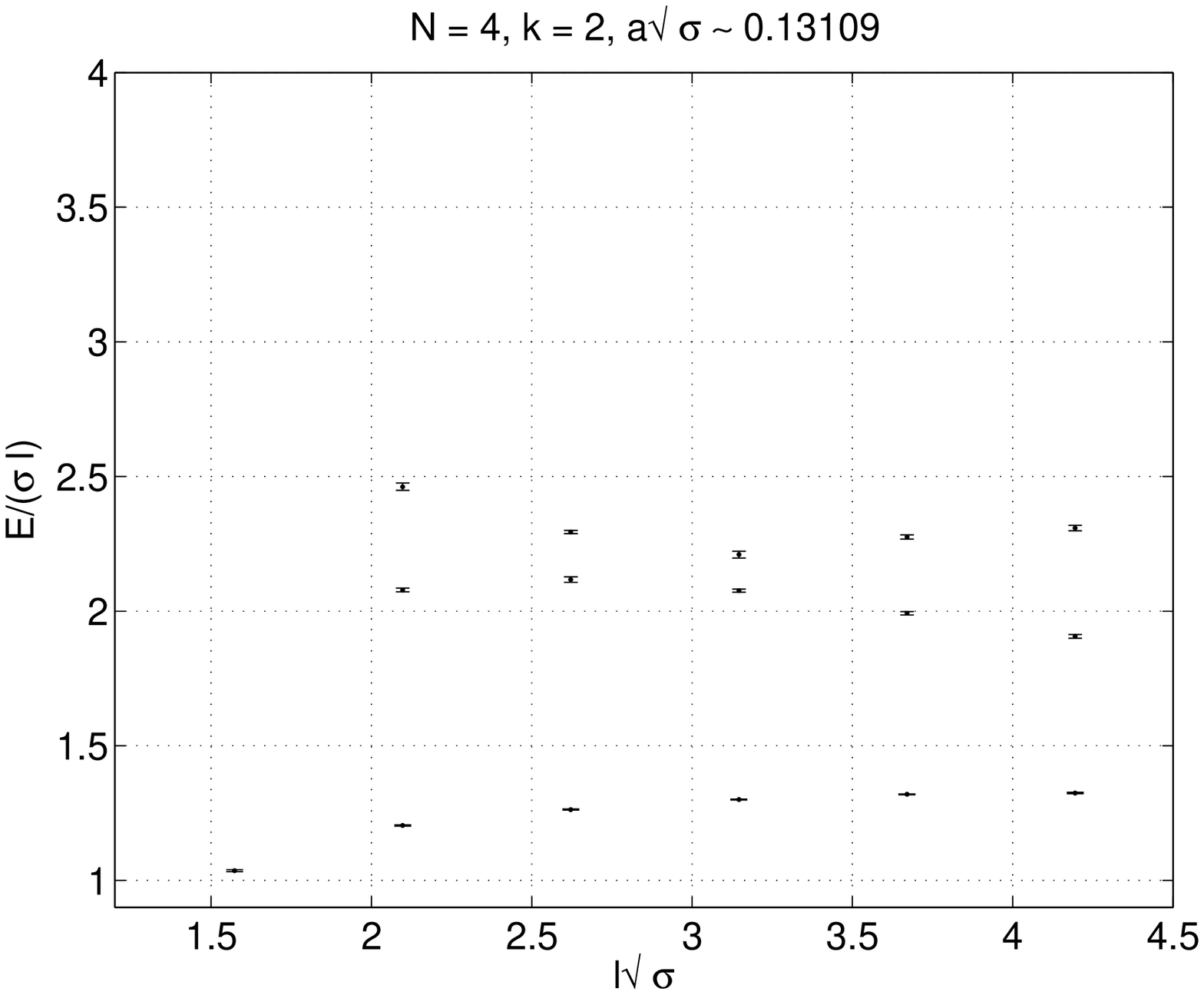} \qquad \includegraphics[width=7cm]{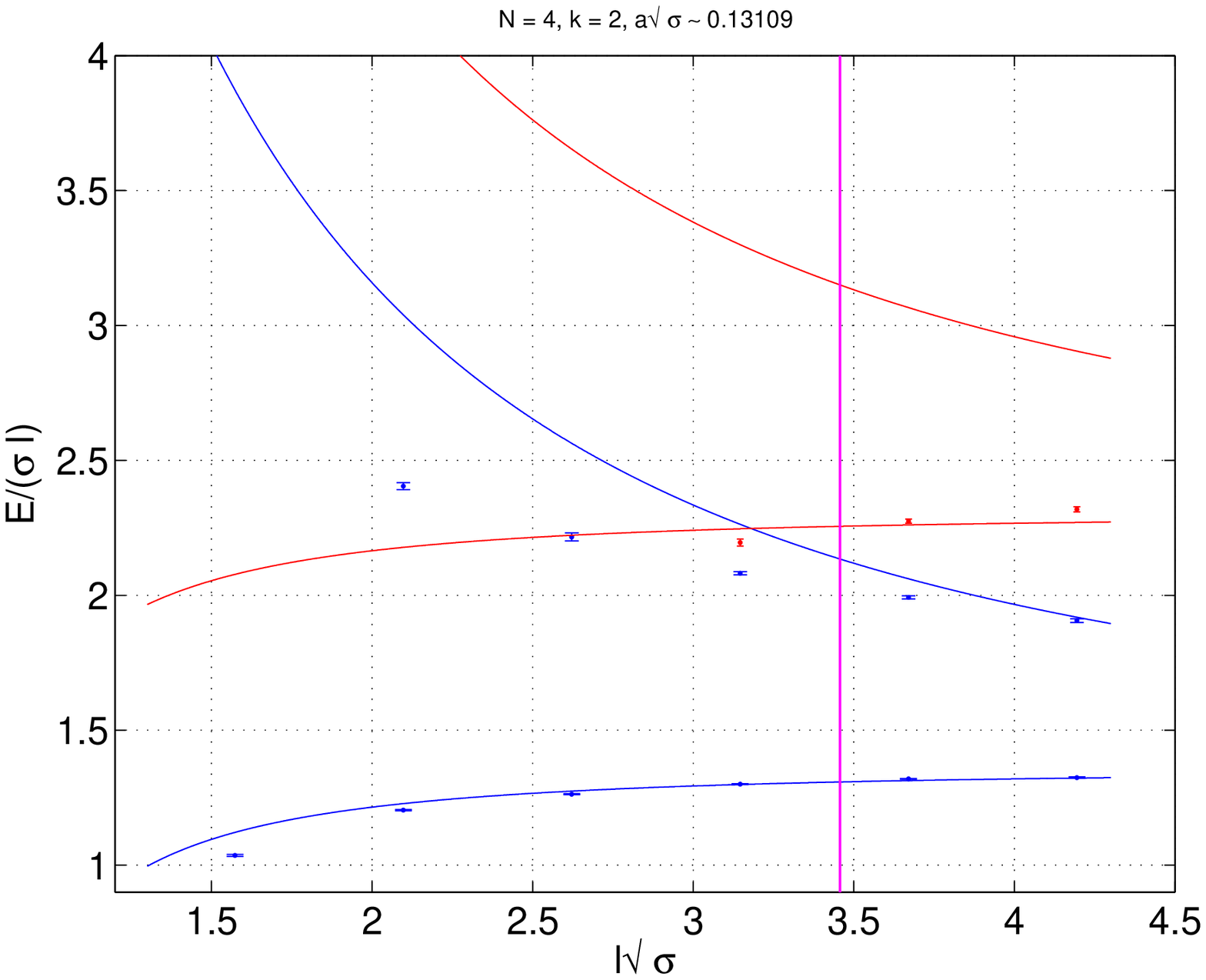} 
} \caption{ \underline{Left:} The energies of the three lowest 
states in the $k=2$ spectrum of $SU(4)$ at $\beta=50.00$ ($a\simeq 
0.06$ fm) vs. the string length. \underline{Right:} Same as in the 
left panel, but obtained after the projection onto symmetric (in 
red) and antisymmetric (in blue) representations.  The lines are 
the predictions of the $2$-NG model (see text), and the vertical line in 
magenta denotes the string length analyzed in \cite{AL}.} 
\label{specN4} 
\end{figure}

Finally, note that whenever two levels cross there appears an approximate degeneracy in the spectrum, which we argue to be the one observed in \cite{LTnLTW,AL}. To check this, we looked at the measured  energies. Performing the variational calculation in the full basis gives the spectrum that we show in the left panel of Fig.~\ref{specN4}.
Performing the variational calculation with only a subspace of operators that belongs to a single representation, we get the spectra in the right panel of Fig.~\ref{specN4}, where we also plot the prediction of a simple 2-NG model for the lowest two states\footnote{The string tensions were chosen to fit our data.}. It is now clear that this model works quite well. Finally, the magenta vertical line denotes the length of the strings analyzed in \cite{AL} and we stress its proximity to the accidental degeneracy point at $l\simeq 3/\sqrt{\sigma}$.
\vspace{-0.6cm}
\section{Summary and future prospects}
\label{summary} 
\vspace{-0.4cm}
We have calculated energies of closed $k$-string in $SU(N)$ gauge theories in 
$2+1$ dimensions. We find that provided the 
strings are longer than $\sim 1.4$ fm, then the deviations of our 
data from the Nambu-Goto (NG) free bosonic string are at most at
the level of $0.5\%$. (This is in contrast to recent results
\cite{Gliozzi} obtained in the $Z_4$ theory). For shorter strings we see significant deviations, which we fit and this allows us to control the systematic error involved in neglecting the $O(1/l^3)$ corrections that are sub leading to the
Luscher term. Doing so, we extract 
tensions from the string energies from a range of lattice spacings and extrapolate the ratio $r_k\equiv \sigma_k/\sigma$ to the continuum. We find that $r_k$ is $1\%-4\%$ higher than the Casimir 
scaling law. We test different large-$N$ extrapolations for 
$r_{2}$ and $r_{N/2}$ and in both cases find that our data naturally prefers a leading $1/N$ correction.
Finally, a striking observation about the spectrum of the excited states is that they fall into separate sectors that correspond to irreducible representations of $SU(N)$. This demonstrates that the string spectrum contains information on the states' $SU(N)$ representation, that goes beyond their N-ality.
 
We stress here that the the results presented in this contribution do not enjoy the same level of
confidence as our former $k=1$ study \cite{KN_papers}, since there are 
several systematic errors that we did not control. The first is the effect of contamination from excited states on the energy estimates obtained from the correlation functions. For $k=1$ we controlled these by performing double-cosh fits to our correlations and saw a shift downwards of $\sim 1-2$ standard deviations, away from the KKN prediction \cite{KN_papers}. For $k>1$ we expect larger contamination from excited states which may push $r_k$ {\em toward} Casimir scaling. Work is now in progress to check for the size of this shift.

Other systematic errors that we currently investigate include the  ($k$-string)/($k$-anti-string) mixing, and the fact that the untraced {\em smeared} Polyakov loops are only approximately $SU(N)$ matrices. Treating these issues may improve the overlap of our operators onto the physical states, and will tighten our control on the classification of the string states according to $SU(N)$ representations.

\end{document}